\documentclass[main]{ssh}
\usepackage{amsmath,bm,amsfonts,algpseudocode,algorithm, textcomp}
\usepackage[dvipsnames]{xcolor}
\DeclareMathOperator*{\argminop}{arg\,min}

\title{Autonomous Electron Tomography Reconstruction with Machine Learning}

\author[1]{William~Millsaps}
\author[2,*]{Jonathan~Schwartz}
\author[3]{Zichao~Wendy~Di}
\author[4]{Yi~Jiang}
\author[2,5,**]{Robert~Hovden}

\affil[1]{Department of Nuclear Engineering $\&$ Radiological Sciences, University of Michigan, Ann Arbor, MI}
\affil[2]{Department of Materials Science and Engineering, University of Michigan, Ann Arbor, MI}
\affil[3]{Mathematics and Computer Science Division, Argonne National Laboratory, Lemont, IL}
\affil[4]{Advanced Photon Source Facility, Argonne National Laboratory, Lemont, IL}
\affil[5]{Applied Physics Program, University of Michigan, Ann Arbor, MI}
\affil[*]{e-mail: jtschw@umich.edu}
\affil[**]{e-mail: hovden@umich.edu}
\date{\today}

\abstract{This manuscript is published in Microscopy and Microanalysis (DOI: 10.1093/micmic/ozad083).

Modern electron tomography has progressed to higher resolution at lower doses by leveraging compressed sensing methods that minimize total variation (TV). However, these sparsity-emphasized reconstruction algorithms introduce tunable parameters that greatly influence the reconstruction quality. Here, Pareto front analysis shows that high-quality tomograms are reproducibly achieved when TV minimization is heavily weighted. However, in excess, compressed sensing tomography creates overly smoothed 3D reconstructions. Adding momentum to the gradient descent during reconstruction reduces the risk of over-smoothing and better ensures that compressed sensing is well behaved. For simulated data, the tedious process of tomography parameter selection is efficiently solved using Bayesian optimization with Gaussian processes. In combination, Bayesian optimization with momentum-based compressed sensing greatly reduces the required compute time---an 80\% reduction was observed for the 3D reconstruction of SrTiO$_3$ nanocubes. Automated parameter selection is necessary for large scale tomographic simulations that enable the 3D characterization of a wider range of inorganic and biological materials.
}

\begin{document}

\maketitle

\section*{Introduction}

For electron tomography, the challenge is to obtain 3D reconstructions that are faithful to the true morphology of inorganic and biological materials. Electron tomography estimates the 3D structure down to sub-nanometer resolution using specimen projections collected in a (scanning) transmission electron microscope (S/TEM) (De Rosier \& Klug, 1968; Hoppe et al., 1974; Midgley \& Weyland, 2003; Scott et al., 2012). Ideally, many projections finely sampled along a $\pm$90$^{\circ}$ angular range would contain high signal-to-noise ratios (SNR). However experimentally, electron tomography is constrained by an insufficient number of projections with mixed Poisson--Gaussian noise (Xu et al., 2015) due to hardware restrictions (Klug \& Crowther, 1972; Schwartz et al., 2022), contamination, and radiation damage (Egerton et al., 2004).

Recent advancements in tomographic reconstruction (Levin et al., 2016) utilize compressed sensing (CS) (Candes et al., 2006; Donoho, 2006) methods that maximize sparsity in the gradient domain. Colloquially called compressed sensing tomography, CS produces superior 3D reconstructions when specimen projections and dose are limited (Goris et al., 2012; Leary et al., 2013; Jiang et al., 2018). However, the benefits of CS tomography are curbed by (1) substantial computation time and (2) sensitivity to tunable hyperparameters; incorrect hyperparameters produce reconstructions that are overly noisy or smoothed, while the proper choice is robust to artifacts (Jiang et al., 2018). Correct selection of hyperparameters requires many reconstructions and hours of compute time to manually evaluate and discover the optimal reconstruction. 

Typically domain experts explore tens to hundreds of hand-selected parameters before identifying a reconstruction most representative of the true specimen structure; failure results in artifacts that may lead to false conclusions of nanostructure (e.g. erroneous pores, incorrect nanoparticle connective structure, etc.). This hand-selection process is particularly problematic and time-consuming when scaling CS tomography across many datasets---as required for simulations that map the effects of different experimental conditions (SNR, sampling, dose, etc.). Fortunately, machine learning provides new opportunities for autonomous optimization of CS tomography. In particular, Bayesian optimization (BO) with Gaussian processes (GP) is designed to identify the global optima of unknown functions that are expensive to evaluate (Rasmussen \& Williams, 2006). BO is an elegant machine learning process that solves regression problems with small training data set sizes while also providing a measure of uncertainty for the predictions. BO has a growing popularity for applications in autonomously driven experiments (Duris et al., 2020; Ziatdinov et al., 2022; Roccapriore et al., 2022), processing of microscopy datasets (Zhang et al., 2021) and experimental design (Deshwal et al., 2021; Liang et al., 2021).

Here we present a framework for autonomous parameter tuning of 3D electron tomography by leveraging Bayesian optimization with Gaussian processes. Using simulated data, we show BO can autonomously and efficiently select optimal hyperparameters for advanced tomography reconstruction algorithms. This not only improves the quality of electron tomograms but also reduces human error associated with compressed sensing. BO for electron tomography simulations reduces parameter exploration here by $\sim$50\% resulting in optimal decisions within 20 test points. In addition, we highlight the benefits of integrating Nesterov momentum into compressed sensing electron tomography, which accelerates the accuracy and rate of convergence. In total, we show the compute time can be reduced by roughly 80\%, and the parameter selection process is entirely automated. This enables a framework for large-scale analysis and validation of novel reconstruction algorithms for electron tomography by efficiently and automatically simulating performance across parameter space.

\section*{Results}
\subsection*{Principles of Reconstruction}

Compressed sensing tomography recovers the 3D structure of specimens by solving an optimization problem that finds a solution which (1) strongly correlates with the experimental projections [e.g. high-angle annular dark-field (HAADF)] and (2) is maximally sparse in the gradient magnitude domain (i.e. minimal total variation, TV$_{\text{min}}$). A common formulation seeks a tomographic reconstruction that minimizes the following unbounded cost function:
\begin{align}
\label{eq:costFunc}
     \argminop_{\bm{x} \geq 0} \quad &\frac{1}{2} \big\| \bm{Ax}-\bm{b}\big \|_2^2 + \lambda \|\bm{x}\|_{\mathrm{TV}}
\end{align} 

where $\bm{b}$ is the measured HAADF, $\bm{x}$ is the reconstructed object, $\bm{A}$ is the measurement matrix, and $\lambda$ is a tunable parameter that balances the two terms (Goris et al., 2012; Leary et al., 2013). The first term denotes the Euclidean distance (occasionally referred to as data distance or DD) between the re-projections and experimental measurements. The second term is the object's total variation (TV), a metric for measuring sparsity in the gradient domain---often referred to as sparsity regularization. Equation~\ref{eq:costFunc} is a non-constrained formulation of the compressed sensing objective. In the constrained formulation, either term is replaced with an equality that bounds the metric within a specified tolerance (Sidky et al., 2011). In either case, the final reconstruction quality is sensitive to selection of a tunable parameter that expresses the balance between maximal sparsity in the gradient domain and agreement with original measurements. Compressed sensing tomography is an optimization problem solvable using several approaches, including simple gradient descent. In this work, we use the fast iterative shrinkage-thresholding algorithm (FISTA) which combines a proximal forward-backward iterative scheme with momentum to obtain reliable reconstructions in accelerated convergence time.

\begin{figure}[ht!]
    \centering
    \includegraphics[width=\columnwidth]{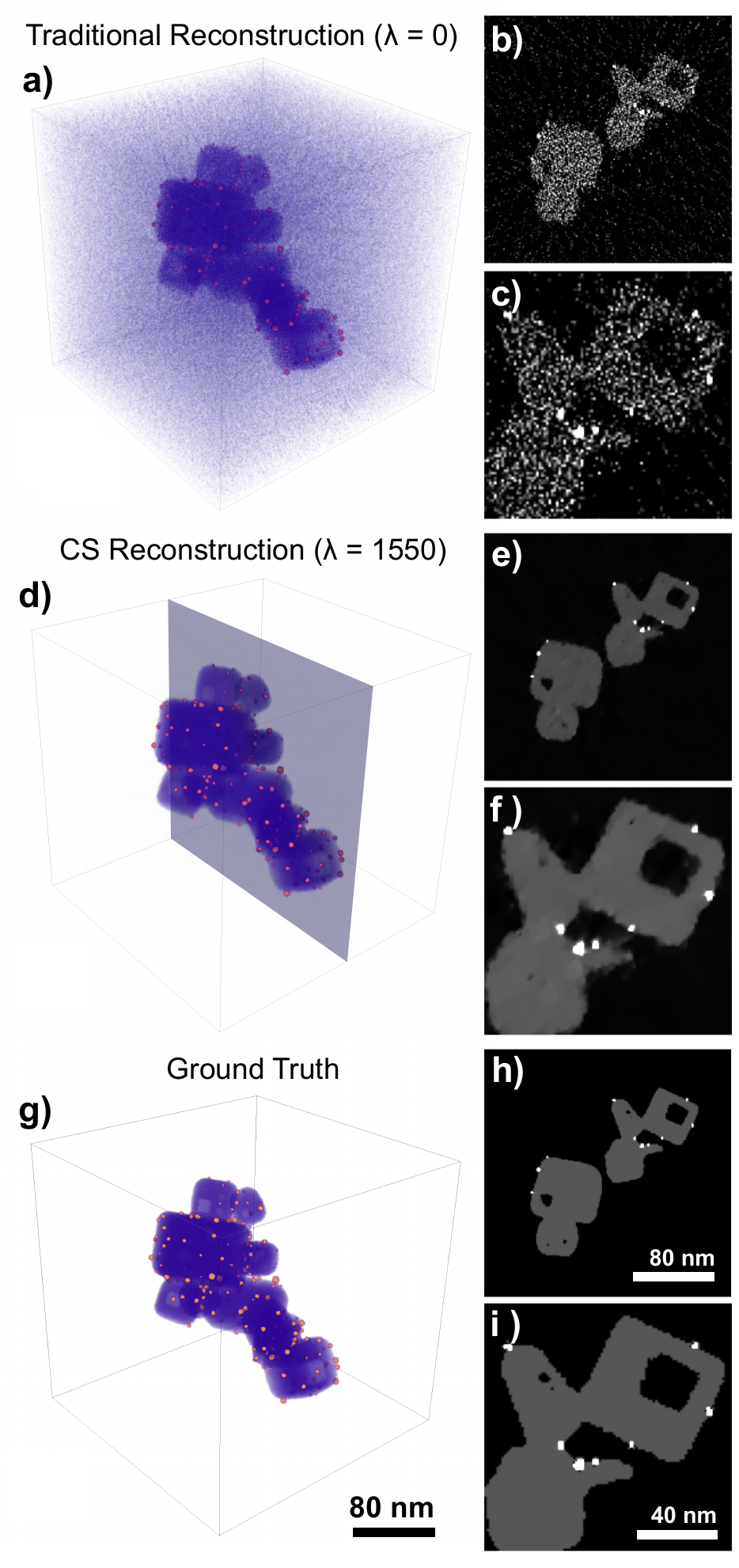}
    \caption{\textbf{Traditional vs compressed sensing electron tomography.} \textbf{a} A traditional reconstruction of 3D strontium titanate (Au / SrTiO$_3$) nanoparticles that relies entirely on minimization of the data distance metric. \textbf{b-c} 2D slices from the reconstruction emphasizing the consequence of traditional algorithms. \textbf{d-f} The optimal CS reconstruction minimizes noise and preserves sharp edges. \textbf{g-i} The ground truth object shows the true morphology of the nanoparticle. The simulation was performed with 47 projections over a $\pm$70$^{\circ}$ tilt range with a $3^{\circ}$ tilt increment and SNR of 5.39. Visualizations rendered using Tomviz.}
    \label{fig::traditionVSoptimal}
\end{figure}

\begin{figure*}[ht!]
    \centering
    \includegraphics[width=\linewidth]{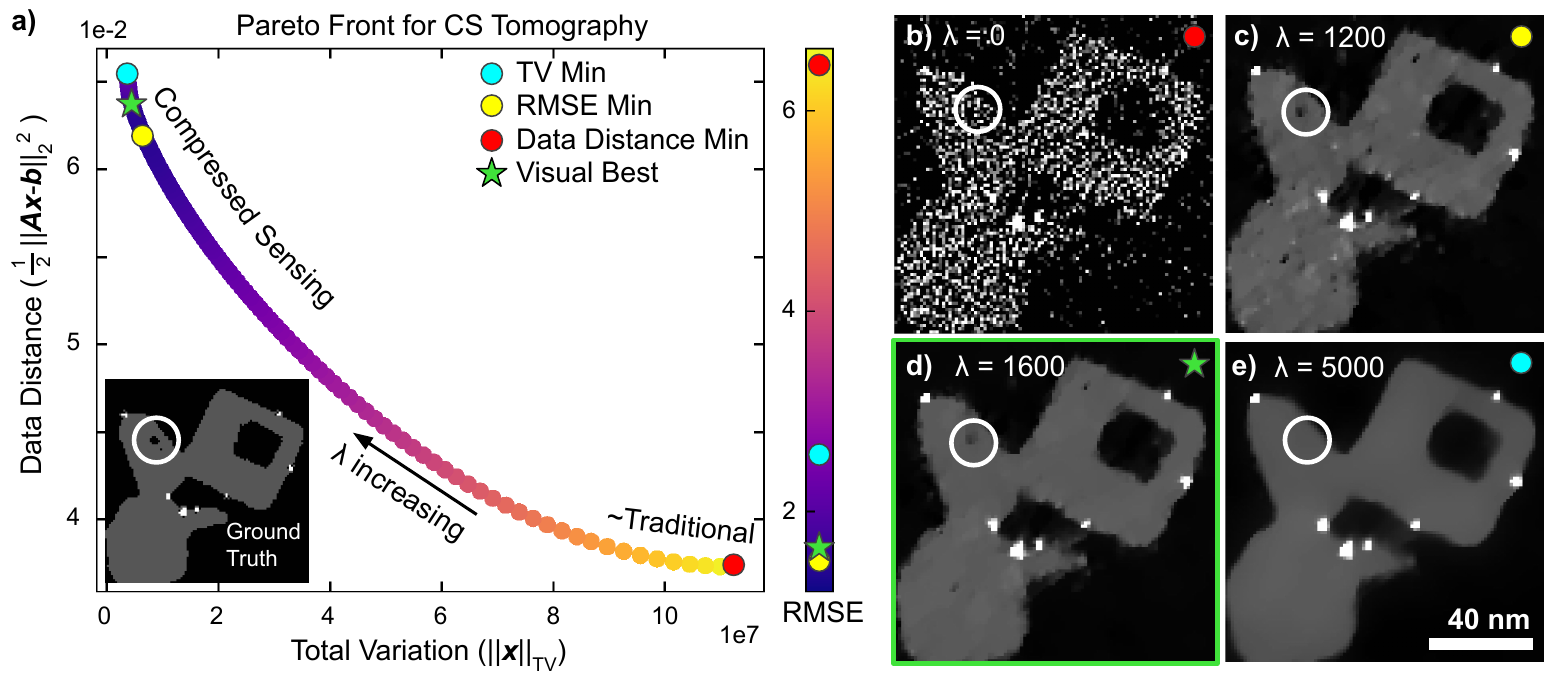}
    \caption{\textbf{Exploring the relationship between tomographic reconstruction quality and $\lambda$ parameter with Pareto fronts.  a} Depicted are the tradeoffs from three reconstruction evaluation metrics: data distance, total variation, and root mean square error. The ground truth (bottom left) is the true morphology of the nanoparticle. RMSE is portrayed through the use of the colorbar. The region of high TV represents traditional tomography, while low TV occurs with compressed sensing. \textbf{b-e} The minima of three metrics, TV, DD, and RMSE, are shown alongside their visual tradeoffs. The white circle highlights a fine detail pore in the object morphology.}
    \label{fig::pareto}
\end{figure*}

\begin{figure*}[ht!]
    \centering
    \includegraphics[width=\linewidth]{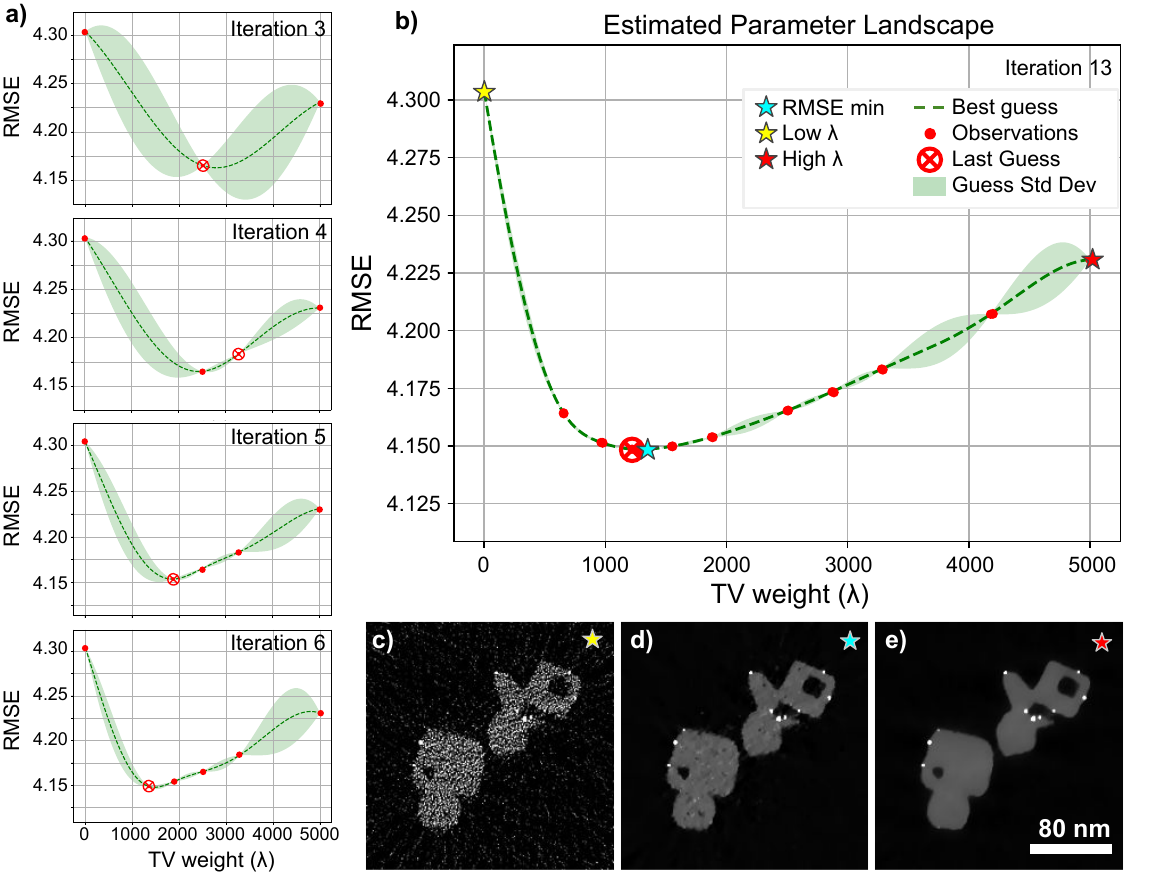}
    \caption{\textbf{Determining the optimal $\lambda$ parameter for simulated compressed sensing electron tomography with Bayesian optimization.} \textbf{a} The BO optimizer iteratively samples $\lambda$ values and performs a tomographic reconstruction to estimate the RMSE landscape. BO strategically determines the next point to sample after each measurement. \textbf{b} The final estimated landscape of RSME vs $\lambda$. The blue star represents the Bayesian optimization's minimum RMSE solution, and the yellow and red represent the minima for the DD and TV metrics, respectively. \textbf{c-e} Corresponding reconstructions: \textbf{c} lies in the traditional regime, \textbf{d} is the RMSE minimizing reconstruction, and \textbf{e} is an over-smoothed reconstruction.}
    \label{fig::Bayesian}
\end{figure*}

\begin{figure*}[ht!]
    \centering
    \includegraphics[width=\linewidth]{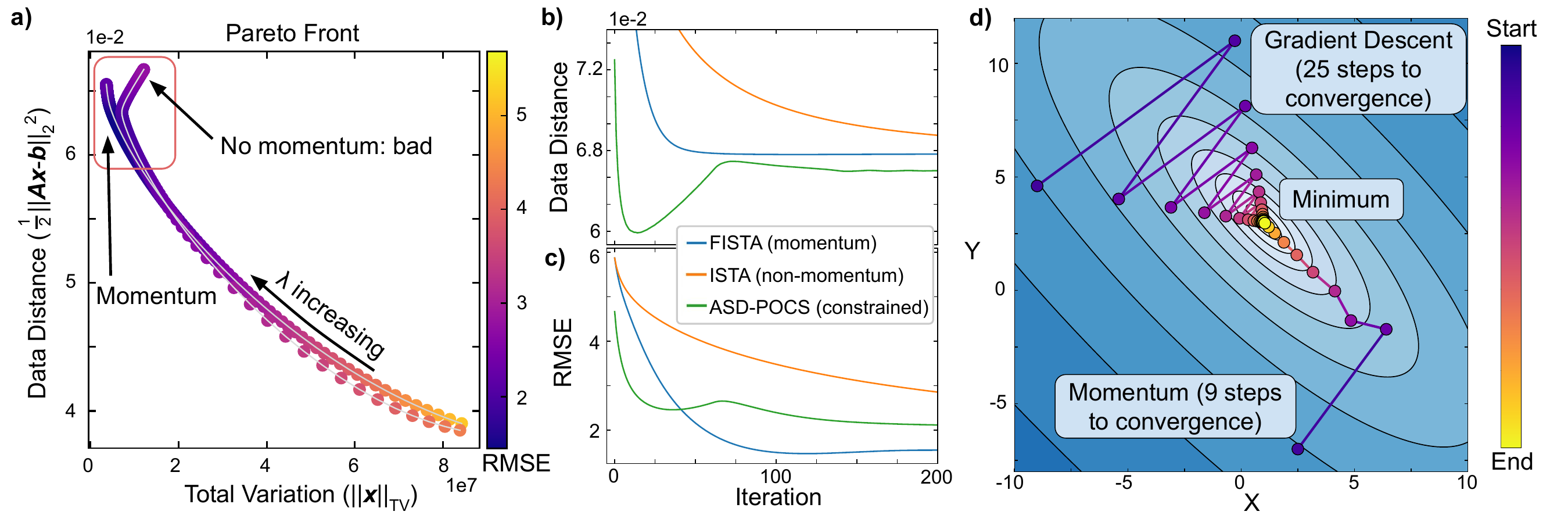}
    \caption{
    \textbf{Improving reconstruction accuracy and convergence speeds with momentum. a} The Pareto front contrasts convergence properties for an algorithm with and without momentum. Without momentum, solutions fail to balance the trade-off between objectives and are not included in the Pareto front. The use of momentum yields solutions that achieve a competitive trade-off of cost metrics. \textbf{b-c} The convergence plots of comparable reconstructions generated by the fast iterative shrinkage-thresholding algorithm (FISTA) implementation (momentum), the iterative shrinkage-thresholding algorithm (ISTA) implementation, and the adaptive steepest descent using projection onto convex sets (ASD-POCS) implementation. FISTA achieves the swiftest convergence (by $>50$ steps) and the best accuracy ($\sim60\%$ better than non-momentum). \textbf{d} Traditional gradient descent (GD) juxtaposed with Nesterov accelerated gradient descent (NAG). From comparable starting points, NAG converges $\sim3$ times faster, requiring 9 steps to reach the minimum, while GD required 25 steps.
    }
    \label{fig::momentum}
\end{figure*}

Figure \ref{fig::traditionVSoptimal} illustrates the superior electron tomography quality offered by compressed sensing for recovery of 3D strontium titanate (Au / SrTiO$_3$) nanoparticles (Padgett et al., 2017). 47 projections simulated over a tilt range of $\pm$70$^{\circ}$ with a 3$^{\circ}$ tilt increment were corrupted with Poisson noise resulting in an average signal-to-noise ratio (SNR) of 5.39 for all of the tilt projections. Conventional iterative algorithms such as the simultaneous iterative reconstruction technique (SIRT (Gilbert, 1972)) or the algebraic reconstruction technique (ART (Gordon et al., 1970)) retain experimental noise and produce streak artifacts when the specimen is undersampled. This class of traditional algorithms only minimizes the data distance (DD) term (Eq. \ref{eq:costFunc}) which suffers from the underdetermined measurement of the 3D specimen. Figure \ref{fig::traditionVSoptimal}a depicts a reconstruction using SIRT. The grainy texture in Figure \ref{fig::traditionVSoptimal}a is primarily an artifact arising from a finite, limited number of HAADF projections as well as thermal and shot noise. Fortunately, compressed sensing algorithms (Beck \& Teboulle, 2009a; Banjak et al., 2018) faithfully recover the gold-decorated nanocube with features nearly identical to the actual structure (Fig. \ref{fig::traditionVSoptimal}d). The improvement in reconstruction quality provided by compressed sensing (TV$_{\text{min}}$) over conventional iterative methods (e.g. SIRT) is visible in both the 3D visualization and 2D slices. In particular, CS avoids the degradation of noise while faithfully preserving the specimen's 5 nm internal pores and cavities. Even in undesirable experimental conditions, accurate material recovery is still possible.

While Figure \ref{fig::traditionVSoptimal} demonstrates compressed sensing under optimal parameter values, significant guess-and-check was performed to validate this solution. Incorrect parameter selection (i.e. the $\lambda$ value) can produce solutions that either lose fine resolution structure or retain noise similar to SIRT reconstructions.  Conventionally, $\lambda$ is hand-tuned by domain expert scientists; to precisely locate minima, the parameter space often requires the exploration of tens to hundreds of reconstructions that are expensive to evaluate (Zhang \& Sonke, 2020). Even with the assistance of GPU acceleration, one reconstruction could take minutes or hours to complete. For reference, a single $256^{3}$ optimized reconstruction takes 10 minutes on a 3840 core GPU (NVIDIA Titan Xp). 3D reconstructions were visualized using the Tomviz software (tomviz.org (Schwartz et al., 2022)). To linearly sample the domain requires 30+ observation points requiring several hours to days, and does not guarantee that the best solution is identified. For this reason, the compute time required for accurate compressed sensing electron tomography is often under-reported or not adequately validated.

\subsection*{Tunable Parameters for Compressed Sensing Tomography}

Solutions produced by compressed sensing electron tomography are significantly influenced by the tunable parameter, $\lambda$, which weights the balance between data consistency and maximal sparsity (TV$_{\text{min}}$). In Figure \ref{fig::pareto}c note the prevalence of erroneous pores and variation in the surface, and in Figure \ref{fig::pareto}e, the absence of expected pores; these artifacts are the consequence of incorrect hyperparameter selection. To best assess reconstruction performance, a Pareto front (Fig. \ref{fig::pareto}a) is generated from many reconstructions across a comprehensive range of $\lambda$ values and the resultant metrics (DD, TV, and RMSE) are plotted. Here, the resulting total variation (TV) and data distance (DD) are the x and y coordinates, respectively, and each point is assigned a colorvalue based on the root mean square error (RMSE).

The Pareto front in Figure \ref{fig::pareto} displays the trade-off between TV minimization and data consistency in our objective function (Eq. \ref{eq:costFunc}) as $\lambda$ is adjusted. Over 500 reconstructions were generated across parameter values ranging from $\lambda = 0$ to $5000$ and plotted based on their final data distance and total variation values. The darker-colored regions on the Pareto front correspond to the favorable, low RMSE (Fig. \ref{fig::pareto}a). While Pareto fronts can be generated from experimental data, calculating the RMSE requires simulated data where the ground truth is known (Schwartz et al., 2020). In general, optimal reconstructions are found in regions with maximal curvature (Tehrani et al., 2012).

A low $\lambda$ (Fig. \ref{fig::pareto}b) produces reconstructions comparable to traditional iterative methods that minimize the data distance with little regard for TV minimization. In the low $\lambda$ regime, reconstructions are degraded by artifacts arising from experimental limitations (e.g. shot noise and sampling), and RMSE is the highest. This corresponds to the bottom right region of the Pareto front where data distance is minimal but TV is high (Fig. \ref{fig::pareto}a).

A high $\lambda$ emphasizes sparsity by minimizing the reconstruction's TV (Fig. \ref{fig::pareto}e). This compressed sensing approach corresponds to the top left region of the Pareto front where the data distance is marginally higher but TV is minimal (Fig. \ref{fig::pareto}a). The RMSE is notably reduced as $\lambda$ is increased (TV minimized). However, $\lambda$ cannot be increased indefinitely as accuracy eventually decreases. Note, the minimum RMSE does not correspond to the minimum TV (leftmost point in the Pareto front), however they are close. This suggests higher $\lambda$ is favorable; but one can overshoot. When $\lambda$ is too high the reconstruction becomes over-smoothed and we lose sufficient resolution to resolve 5 nm pores in Figure~\ref{fig::pareto}e. Between the minimal RMSE (Fig. \ref{fig::pareto}c) and minimal TV (Fig. \ref{fig::pareto}e)) reconstructions, there is a ``visual best'' reconstruction that manages the trade-off of sharpness and feature preservation. The visual best retains accurate information about the true morphology as highlighted by the presence of pores (white circle), while also minimizing the presence of inaccurate noise-like features.
 
Although Pareto fronts provide valuable assessment of compressed sensing tomography, the large number of reconstructions required is computationally expensive. An efficient approach to selecting the optimal $\lambda$ parameter is needed.

\subsection*{Autonomous Reconstruction with Bayesian Optimization}

Bayesian optimization (BO) with Gaussian processes (GPs) is designed to identify the global optima of unknown functions and is well suited for reliably identifying preferred $\lambda$ values in compressed sensing electron tomography (Mockus, 1989). Here we show BO with GPs enables the autonomous optimization of tomography parameters for simulated datasets and identifies the optimal solution in under 20 reconstructions. We found this improvement in parameter exploration typically results in 50\% time reduction without the need for human supervision.  

Bayesian optimization involves two core components: (1) develop a posterior probability distribution of the parameter objective with GP regression and (2) manage the recommendation of future parameter observations using an exploration function---commonly referred to as an acquisition function (Jonas, 1994). This `exploration function' is the procedure for deciding the next parameter to evaluate. BO uses GPs to model the objective as a joint distribution of functions whose landscape (e.g. smoothness) is defined by a kernel or covariance function. In addition to modeling the objective, the posterior distribution quantifies the uncertainty (standard deviation) over all unsampled points. The exploration function is given the current knowledge of the system (known data points), a fitting strategy (GP), and an exploration strategy. The exploration function proposes sampling points while balancing the importance between exploitation of extrema and exploration in regions with large uncertainty. Both phases act as a safeguard against convergence to local optimum by exploring the entire parameter space (Cao et al., 2022).

Figure \ref{fig::Bayesian} shows the Bayesian optimization process to systematically determine the optimal $\lambda$ value from estimates of the RMSE landscape. As more points are gathered, certainty of the RMSE landscape improves. The variance in its estimate of the landscape is used to predict the global minimum. The estimated landscape is updated after each prediction until a certain number of iterations is performed or the solution converges. It is critical that the global minimizer is within the bounds provided to the optimizer. Fig. \ref{fig::Bayesian}b shows the final landscape estimate. For the SrTiO$_3$ nanoparticle dataset, Bayesian optimization confidently identifies the solution with minimal error within approximately $13$ iterations (Fig. \ref{fig::Bayesian}b).  Figures \ref{fig::Bayesian}c-e display nanoparticle reconstructions from points on the landscape. Empirically, we find BO consistently discovers the optimal parameter within 12 to 20 estimates on datasets tested using compressed sensing (TV$_{\text{min}}$) reconstruction. As shown by Fig. \ref{fig::Bayesian}b, the relationship between $\lambda$ and RMSE is convex, with a well-defined global minimum that allows for deterministic optimal solutions.

\subsection*{Accelerating Convergence and Accuracy with Momentum}
Incorporating momentum acceleration into compressed sensing electron tomography prevents heavily over-smoothed solutions and provides faster reconstruction. At a high level, momentum leverages previous steps during reconstruction as prior information for the current step to provide a multiplicative boost in convergence. We can include momentum in standard gradient descent optimization through the addition of an extrapolation step after the gradient update:
\begin{align}
\label{eq:momentum}
     \bm{x}_{t} &= \bm{y}_{t-1} - \alpha \nabla f(\bm{y}_{t-1}) \\
     \bm{y}_{t} &= \bm{x}_{t} + \beta_t (\bm{x}_{t} - \bm{x}_{t-1})
\end{align} 
where $\bm{x}_t$ is the current iterate, $\alpha \nabla f(\bm{y}_{t-1})$ is the gradient update scaled by a step size $\alpha$, and $\beta_t(\bm{x}_{t} - \bm{x}_{t-1})$ is the momentum acceleration scaled by $\beta_t$, which can vary with $t$ (Polyak, 1964).
Conceptually the momentum term nudges each iteration further down the parabolic landscape by descending further along the previous update's direction and dampens potential oscillations. While the gradient descent algorithm is simple and guarantees monotonic improvement for every iteration, the fixed learning rate can slowly progress toward the local solution. Increasing the learning poses the risk of oscillating in regions of high curvature. Momentum-based descent adaptively scales the step size in response to the local curvature per iteration. The advantage of momentum-based optimization can be seen in the fast iterative shrinkage-thresholding algorithm (FISTA) (Beck \& Teboulle, 2009a)---one example of compressed sensing that minimizes the sum of a convex data fidelity term and a non-smooth penalty term (e.g. TV). This simple modification remarkably achieves an optimal quadratic convergence rate (Beck \& Teboulle, 2009a). Specifically, Nesterov has shown analytically that the convergence rate of these classic first-order methods can be sped up from $\mathcal{O}(1/n)$ to $\mathcal{O}(1/n^2)$ after $n$ iterations with the incorporation of momentum (Nesterov, 1983).

The Pareto fronts in Figure \ref{fig::momentum}a compare non-momentum and momentum-based compressed sensing tomography and are nearly identical but deviate in the regime where $\lambda$ values are high (top-left). In particular, we see simple gradient descent fails if $\lambda$ is too large as the Pareto optima diverge into a regime where TV, DD, and RMSE all increase. This is expected since the classical convergence bounds for simple gradient descent makes assumptions on the cost function smoothness such as requiring a Lipschitz-continuous gradient (Shamir \& Zhang, 2013). In contrast, momentum-based optimization converges to viable solutions with minimal TV---comparable to Figure \ref{fig::pareto}a. Thus, adding a momentum term places a ceiling on the amount of error one can obtain from an incorrect $\lambda$ selection.

Momentum-based compressed sensing also offers faster convergence, meaning each 3D reconstruction completes in fewer iterations and shorter compute time. The RMSE and DD convergence for momentum-based FISTA, the non-momentum iterative shrinkage-thresholding implementation (ISTA (Daubechies et al., 2004)), and a popular constrained CS algorithm (ASD-POCS (Sidky \& Pan, 2008)) are quantitatively shown in Figure \ref{fig::momentum}b-c. Here, FISTA achieves the greatest reconstruction accuracy and the swiftest convergence, requiring only 125 iterations while non-accelerated implementations need between 200 and 300 iterations to converge. Faster convergence is a known benefit of adding momentum (Xu et al., 2016) and can be illustrated by seeking the global minimum of a synthetic landscape known as Booth's Function (Fig. \ref{fig::momentum}d)---a convex function often used to test optimization performance (Jamil \& Yang, 2013). For the Booth landscape demonstration, standard gradient descent requires roughly 25 iterations to reach the minimum due to a slowdown near the flat minimum. With momentum, the descent retains speed and only requires $\sim40\%$ of the steps to reach the solution (9 in total).

\section*{Conclusions}

Recent progress in electron tomography achieves higher resolution at lower doses by leveraging compressed sensing methods that minimize total variation. However, these optimization-based reconstruction algorithms introduce tunable parameters that greatly influence the reconstruction quality. This work presents various methods to reliably tune hyperparameters with Pareto fronts, Bayesian optimization, and momentum-based gradient descent. Pareto fronts provide intuitive assessment of the reconstruction accuracy across parameter space. For simulated data, autonomous parameter tuning for compressed sensing electron tomography is achievable using Bayesian optimization with Gaussian processes. BO with GPs autonomously reduces the computational time required to discover optimal tomography parameters by 50\% without any user intervention. This framework can be extended to experimental datasets, however, additional work is required to identify target metrics for optimization other than RMSE. Substitutes for RMSE may include (data distance, total variation, sharpness, or inflection points of a Pareto front). Momentum accelerated gradient descent algorithms can further reduce compute time by roughly 60\% and better guarantee accurate reconstruction when total variation minimization is heavily emphasized. This work is essential for large-scale tomographic simulations that require automatic parameter selection and efficient use of computational resources. Pareto front analysis and momentum-based descent are directly applicable to experimental reconstruction---and should be utilized. However, experimental use of Bayesian optimization requires further work to identify a substitute for calculating RMSE. With a better understanding of compressed sensing electron tomography---how it succeeds and fails---we can recover the 3D structure of a wider range of inorganic and biological materials with higher fidelity and resolution.

\section*{Methods}
\subsection*{Compressed Sensing Electron Tomography using \\
Fast Iterative Shrinkage-Thresholding Algorithm}

Compressed sensing tomography was reconstructed using the fast iterative shrinkage-thresholding algorithm (FISTA) (Beck \& Teboulle, 2009a). FISTA is a proximal gradient algorithm that solves optimization problems of the following form: $\argminop_x~f(x)~+~\mathcal{R}(x)$, where $f$ is a continuously differentiable convex function (i.e. data fidelity with least squares) and $\mathcal{R}$ could be a nonsmooth regularizer (e.g. TV$_{\text{min}}$). A description for solving Eq. \ref{eq:costFunc} with FISTA is given in Algorithm \ref{alg:FistaAlg}: 

\begin{algorithm}
\caption{FISTA for Electron Tomography}
\begin{algorithmic}
\State \textbf{Initialize: } $\bm{x}_0 = \bm{y}_{0} = \bm{0}, t_0 = 1, L_A = \|\bm{A}\|^2_2,~\bar{\lambda} := \lambda/L_A$
\For{$k = 1, N_{iter}$} \Comment{Main Loop}
    \State  $\bm{y}_k = \bm{y}_{k-1} - \frac{1}{L_A} \bm{A}^{\top} (\bm{A}\bm{y}_{k-1} - \bm{b})$ \Comment{Gradient Descent}
    \State  $\bm{x}_k = \argminop_{\bm{y}}~\|\bm{y}\|_{\text{TV}}~+~1/(2\bar{\lambda})||\bm{y}_k-\bm{y}||^2_2 $ \Comment{TV Min}
    \State  $t_{k} = \frac{1}{2}(1+\sqrt{1+4t_{k-1}})$ \Comment{Momentum weighting}
    \State  $\bm{y}_{k+1} = \bm{x}_k + \frac{t_{k-1}-1}{t_{k}}(\bm{x}_k - \bm{x}_{k-1})$ \Comment{Momentum update}
\EndFor
\State \textbf{Return: } $\bm{x}_{\text{Niter}}$ \Comment{Return the Final Reconstruction}
\end{algorithmic}
\label{alg:FistaAlg}
\end{algorithm}
\noindent Where $\bm{b}$ is the measured HAADF, $\bm{A}$ is the measurement matrix, $\bm{x}$ and $\bm{y}$ are reconstructed objects, $\lambda$ is the regularization parameter, $L_A$ is the Lipschitz parameter, and $k$ is the iteration number. 

The algorithm is composed of three key steps. First, gradient descent on the smooth data fidelity function to produce an intermediate reconstruction denoted as $\bm{y}_k$. In the subsequent smoothing step, we evaluate the isotropic TV proximal operator to denoise the reconstructions, which in itself is another iterative algorithm (Beck \& Teboulle, 2009b; Parikh \& Boyd, 2014). The smoothing step can incorporate non-negativity or box constraints. For the non-accelerated IST algorithm, the two associated momentum steps (weighting and updating) are ignored.

\subsection*{Bayesian Optimization with Gaussian Processes}
\noindent A Gaussian process (GP) is a probabilistic tool that can model complex, non-linear relationships in data. It is a generalization of the Gaussian distribution and it is often used as a prior over functions in Bayesian modeling. In a GP, any point in the function's domain (i.e., input space) is associated with a Gaussian distribution over possible function values (i.e., output space). The mean and covariance of these Gaussian distributions are determined by a mean function and a covariance function (also known as a kernel), which are typically chosen based on prior knowledge or domain expertise. We carried out BO in Python with the Scikit Optimize library (scikit-optimize.github.io/stable) with the Matern kernel and GP Hedge acquisition strategy (Brochu et al., 2011). The Matern kernel is a generalization of the squared exponential kernel and is particularly useful for modeling smooth functions. The GP-Hedge acquisition framework considers a portfolio comprising a set of predefined acquisition functions (e.g. lower confidence bound or probability of improvement). Each function provides a candidate to sample at every iteration and GP hedge uses a probability-based gain strategy to determine which function has the optimal recommendation. After fitting the surrogate model to the evaluated point, each function receives a score based on its prediction.

\section*{References}
\footnotesize
\noindent
Banjak, H., Grenier, T., Epicier, T., Koneti, S., Roiban, L., Gay, A.-S., Magnin, I., Peyrin, F., \& Maxim, V. (2018). Evaluation of noise and blur effects with SIRT-FISTA-TV reconstruction algorithm: Application to fast environmental transmission electron tomography. Ultramicroscopy 189, 109-123. \vspace{0pt}

\noindent
Beck, A. \& Teboulle, M. (2009a). A Fast Iterative Shrinkage-Thresholding Algorithm for Linear Inverse Problems. SIAM J. Imaging Sciences 2(1), 183-202.\vspace{0pt}

\noindent
Beck, A. \& Teboulle, M. (2009b). Fast Gradient-Based Algorithms for Constrained Total Variation Image Denoising and Deblurring Problems. IEEE Transactions on Image Processing 18(11), 2419-2434. \vspace{0pt}

\noindent
Brochu, E., Hoffman, M., \& de Freitas, N. (2011). Portfolio Allocation for Bayesian Optimization in Proceedings of the Twenty-Seventh Conference on Uncertainty in Artificial Intelligence, Cozman F, \& Pfeffer A (Eds.), AUAI Press 327-336. \vspace{0pt}

\noindent
Candés, E. J., Romberg, J. \& Tao, T. (2006). Robust uncertainty principles: exact signal reconstruction from highly incomplete frequency information. IEEE Trans. Inf. Theory 52(2), 489-509. \vspace{0pt}

\noindent
Cao, M. C., Chen, Z., Jiang, Y., \& Han, Y. (2022). Automatic parameter selection for electron ptychography via Bayesian optimization. Scientific Reports 12, 12284.\vspace{0pt}

\noindent
Daubechies, I., Defrise, M., \& De Mol, C. (2004). An iterative thresholding algorithm for linear inverse problems with a sparsity constraint. Communications on Pure and Applied Mathematics 57(11), 1413-1457. \vspace{0pt}

\noindent
De Rosier, D. J. \& Klug, A. (1968). A Reconstruction of Three Dimensional Structures from Electron Micrographs. Nature 217, 130-134.\vspace{0pt}

\noindent
Deshwal, A., Simon, C. M. \& Doppa, J. R. (2021). Bayesian optimization of nanoporous materials. Mol. Syst. Des. Eng. 6, 1066-1186.\vspace{0pt}

\noindent
Donoho, D. L. (2006). Compressed Sensing. IEEE Trans. Inf. Theory 52(4), 1289-1306.\vspace{0pt}

\noindent
Duris, J., Kennedy, D., Hanuka,  A., Shtalenkova, J., Edelen, A., Baxevanis, P., Egger, A., Cope, T., McIntire, M., Ermon, S., \& Ratner, D. (2020). Bayesian Optimization of a Free-Electron Laser. Phys. Rev. Lett. 124, 124801. \vspace{0pt}

\noindent
Egerton R.F., Li, P., \& Malac, M. (2004). Radiation damage in the TEM and SEM. Micron 35(6), 399-409. \vspace{0pt}

\noindent
Gilbert, P. (1972). Iterative methods for the three-dimensional reconstruction of an object from projections. J. Theor. Biol. 36(1), 105-117. \vspace{0pt}

\noindent
Gordon, R., Bender, R., \& Herman, G. T. (1970). Algebraic Reconstruction Techniques (ART) for three-dimensional electron microscopy and X-ray photography. J. Theor. Biol. 29(3), 471-481. \vspace{0pt}

\noindent
Goris, B., Van den Broek, W., Batenburg, K.J, Heidari Mezerji, H., \& Bals, S. (2012). Electron tomography based on a total variation minimization reconstruction technique. Ultramicroscopy 113, 120–130. \vspace{0pt}

\noindent
Hoppe, W., Gassmann, J., Hunsmann, J., Schramm H. J., \& Sturm, M. (1974). Three-dimensional reconstruction of individual negatively stained yeast fatty-acid synthetase molecules from tilt series in the electron microscope. Hoppe Seylers Z Physiol  Chem 355(11), 1483-1487.  \vspace{0pt}

\noindent
Jamil, M. \& Yang, X. (2013). A Literature Survey of Benchmark Functions For Global Optimization Problems. International Journal of Mathematical Modelling and Numerical Optimisation 4(2), 150-194.\vspace{0pt}

\noindent
Jiang, Y., Padgett, E., Hovden, R. \& Muller, D. A. (2018). Sampling limits for electron tomography with sparsity-exploiting reconstructions. Ultramicroscopy 186, 94-103. \vspace{0pt}

\noindent
Jonas, M. (1994). Application of Bayesian approach to numerical methods of global and stochastic optimization. Journal of Global Optimization 4, 347-365.\vspace{0pt}

\noindent
Klug, A. \& Crowther, R. A. (1972). Three-dimensional Image Reconstruction from the Viewpoint of information Theory. Nature 238, 435-440.\vspace{0pt}

\noindent
Leary, R., Saghi, Z., Midgley, P. A., \& Holland, D. J. (2013). Compressed sensing electron tomography. Ultramicroscopy 131, 70-91.\vspace{0pt}

\noindent
Levin, B. D. A., Padgett, E., Chen, C.-C., Scott, M. C., Xu, R., Theis, W., Jiang, Y., Yang, Y., Ophus, C., Zhang, H., Ha, D.-H., Wang, D., Yu, Y., Abruña, H. D., Robinson, R. D., Ercius, P., Kourkoutis, L. F., Miao, J., Muller, D. A., \& Hovden, R. (2016). Nanomaterial datasets to advance tomography in scanning transmission electron microscopy. Scientific Data 3, 160041. \vspace{0pt}

\noindent
Liang, Q., Gongora, A. E., Ren, Z., Tiihonen, A., Liu, Z., Sun, S., Deneault, J. R., Bash, D., Mekki-Berrada, F., Khan, S. A., Hippalgaonkar, K., Maruyama, B., Brown, K. A., Fisher III, J. \& Buonassisi, T. (2021). Benchmarking the performance of Bayesian optimization across multiple experimental materials science domains. npj Comput. Mater. 7, 188. \vspace{0pt}

\noindent
Midgley, P. \& Weyland, M. (2003). 3D electron microscopy in the physical sciences: the development of Z-contrast and EFTEM tomography. Ultramicroscopy 96(3-4), 413-431.\vspace{0pt}

\noindent
Mockus, J. (1989). Bayesian Approach to Global
Optimization: Theory and Applications. Dordrecht,The Netherlands: Kluwer
Academic Publishers.\vspace{0pt}

\noindent
Nesterov, Y. E. (1983). A method of solving a convex programming problem with convergence rate O(1/k\textasciicircum2). Soviet Mathematics Doklady 27, 372-376. \vspace{0pt}

\noindent
Padgett, E., Hovden, R., DaSilva, J. C., Levin, B. D. A., Grazul, J. L., Hanrath, T., \& Muller, D. A. (2017). A Simple Preparation Method for Full-Range Electron Tomography of Nanoparticles and Fine Powders. Microscopy and Microanalysis 23(6), 1150-1158. \vspace{0pt}

\noindent
Parikh, N. \& Boyd, S. (2014). Proximal Algorithms. Foundations and Trends in Optimization 1(3), 123-239.\vspace{0pt}

\noindent
Polyak, B. T. (1964). Some methods of speeding up the convergence of iteration methods. USSR Computational Mathematics and Mathematical Physics 4(5), 1-17. \vspace{0pt}

\noindent
Rasmussen, C. E. \& Williams, C. (2006). Gaussian Processes for Machine Learning. Cambridge, Massachusetts: MIT Press. \vspace{0pt}

\noindent
Roccapriore, K. M., Kalinin, S. V., \& Ziatdinov, M. (2022). Physics Discovery in Nanoplasmonic Systems via Autonomous Experiments in Scanning Transmission Electron Microscopy. Advanced Science 9(36), 2203422. \vspace{0pt}

\noindent
Schwartz, J., Zheng, H., Hanwell, M., Jiang, Y., \& Hovden, R. (2020). Dynamic compressed sensing for real-time tomographic reconstruction. Ultramicroscopy 219, 113122. \vspace{0pt}

\noindent
Schwartz, J., Harris, C., Pietryga, J., Zheng, H., Kumar, P., Visheratina, A., Kotov, N. A., Major, B., Avery, P., Ercius, P., Ayachit, U., Geveci, B., Muller, D. A., Genova, A., Jiang, Y., Hanwell, M., \& Hovden, R. (2022). Real-time 3D analysis during electron tomography using tomviz. Nat. Commun. 13, 44-58. \vspace{0pt}

\noindent
Scott, M. C., Chen, C.-C., Mecklenburg, M., Zhu, C., Xu, R., Ercius, P., Dahmen, U., Regan, B. C., \& Miao, J., (2012). Electron tomography at 2.4-angstrom resolution. Nature 483, 444-447. \vspace{0pt}

\noindent
Shamir, O. \& Zhang, T. (2013). Stochastic Gradient Descent for Non-smooth Optimization Convergence Results and Optimal Averaging Schemes in Proceedings of the 30th International Conference on Machine Learning. Proceedings of Machine Learning Research 28(1), 71-79. \vspace{0pt}

\noindent
Sidky, E. Y. \& Pan, X. (2008). Image reconstruction in circular cone-beam computed tomography by constrained, total-variation minimization. Pys. Med. Biol. 53(17), 4777-4807. \vspace{0pt}

\noindent
Sidky, E. Y., Duchin, Y., Pan, X., \& Ullberg, C. (2011). A constrained, total-variation minimization algorithm for low-intensity x-ray CT. Med. Phys. 38(S1), S117-S125. \vspace{0pt}

\noindent
Tehrani, J. N., McEwan, A., Jin, C. \& van Schaik, A. (2012). L1 regularization method in electrical impedance tomography by using the L1-curve (Pareto frontier curve). Applied Mathematical Modelling 36(3), 1095-1105. \vspace{0pt}

\noindent
Xu, R., Chen, C.-C., Wu, L., Scott, M. C., Theis, W., Ophus, C., Bartels, M., Yang, Y., Ramezani-Dakhel, H., Sawaya, M. R., Heinz, H., Marks, L. D., Ercius, P., \& Miao, J. (2015). Three-dimensional coordinates of individual atoms in materials revealed by electron tomography. Nat. Mater. 14, 1099-1103. \vspace{0pt}

\noindent
Xu, Q., Yang, D., Tan, J., Sawatzky, A., \& Anastio, M. A. (2016). Accelerated fast iterative shrinkage thresholding algorithms for sparsity-regularized cone-beam CT image reconstruction. Medical Physics 43(4), 1849-1872. \vspace{0pt}

\noindent
Zhang, H. \& Sonke, J.-J. (2020). Pareto frontier analysis of spatio-temporal total variation based four-dimensional cone-beam CT. Biomedical Physics \& Engineering Express 5(6), 065011. \vspace{0pt}

\noindent
Zhang, C., Baraissov, Z., Duncan, C., Hanuka, Adi., Edelen, A., Maxson, J., \& Muller, D. A. (2021). Aberration Corrector Tuning with Machine-Learning-Based Emittance Measurements and Bayesian Optimization. Microscopy and Microanalysis 27(S1), 810-812. \vspace{0pt}

\noindent
Ziatdinov, M., Liu, Y., Kelley, K., Vasudevan, R., \& Kalini, S. V. (2022). Bayesian Active Learning for Scanning Probe Microscopy: From Gaussian Processes to Hypothesis Learning. ACS Nano 16, 13492-13512. \vspace{0pt}

\section*{Acknowledgements}

W.M. and J.S. acknowledge support from the Dow Chemical Company. R.H. would like to acknowledge support from the DOE Office of Basic Energy Sciences. This research used the Argonne Leadership Computing Facility at Argonne National Laboratory, which is supported by the Office of Science of the U.S. Department of Energy under Contract No. DE-AC02-06CH11357.

\section*{Ethics Declarations}
\subsection*{Competing Interests}
The authors declare no competing interests.

\end{document}